\documentstyle[12pt ,aasms4]{article}
\begin{document}
\baselineskip=18pt 
\def\be{\begin{equation}} 
\def\ee{\end{equation}} 
\def\bearst{\begin{eqnarray*}} 
\def\eearst{\end{eqnarray*}} 
\begin{center}
{\Large 
{\bf A Dynamical Approach to a Self-similar Universe}} 
\vskip 1.1cm 
{\bf E. Abdalla$^a$, N. Afshordi$^b$, 
K. Khodjasteh$^c$ and R. Mohayaee$^d$}%
\\ \vskip 0.3cm 
{$^{a,d}${\it Instituto de F\'\i sica, USP, C.P.66.318, Sao
Paulo, Brazil\\ 
\vskip 0.3cm 
$^{bc}$Physics Department, Sharif University,
P.O. Box 8639, Tehran, Iran\\ 
\vskip 0.3cm $^d$IPM, P.O. Box 19395-5531,
Tehran, Iran\\ 
\vskip 0.3cm 
$^a$eabdalla@fma.if.usp.br 
\hskip 2mm 
$^b$niayesh@rose.ipm.ac.ir \\ 
\vskip 0.3cm 
$^c$khojaste@physic.sharif.ac.ir 
\hskip 2mm 
$^d$mohayaee@theory.ipm.ac.ir }}
\end{center}

\abstract 

We write a non-relativistic Lagrangian for a hierarchical universe. 
The equations of motion are solved numerically and the evolution of 
the fractal dimension is obtained for different initial conditions. 
We show that our model is homogeneous at the time of the last scattering, 
but evolves into a self-similar universe with a remarkably constant 
fractal dimension. We also show that the Hubble law is implied by this 
model and make an estimate for the age of the universe.

\vfill\eject


\section{Introduction}

\indent

This is a decisive time for cosmology since our theories for 
the large scale
structure of the universe are being seriously challenged by 
the ever-growing amount of data.

The CfA1 redshift survey (de Lapparent, Geller \& Huchra, 1986, 1988) was
the first to reveal structures such as filaments and voids on scales where a
random distribution of matter was expected. The most remarkable feature of
these structures is the so-called ``great wall" which is a coherent sheet of
galaxies extended over an area of at least $60\times 170 $ Mpc (Geller \&
Huchra 1989). Later on, deep pencil beam surveys (Broadhurst et al. 1990),
the redshift surveys based on IRAS catalogue (Efstathiou et al. (1990a,
1990b), Saunders et al. 1991, Fisher et al 1996), the deep wide angle survey
SSRS (de Costa 1988, 1994) and some others have shown inhomogeneities at
scales where the galaxy-galaxy and cluster-cluster correlations were
believed to be negligible. Of particular significance for the future are the
two extensive redshift surveys, SLOAN and 2dF, about to commence, which aim
to trace the 3-dimensional distribution of over one million galaxies across
the northern and southern skies.

The first quantitative study of the cosmic inhomogeneity lead to the
well-known $1.8$ power law behaviour of the galaxy-galaxy correlation
function (Groth \& Peebles 1977, Peebles 1980, Davis \& Peebles (1983a,
1983b)). Although this law has been consistently identified in different
catalogues, the break away from it at larger scales and a crossover to
homogeneity has not yet been established (Davis 1996, Pietronero 1987, 
Coleman \& Pietronero 1992, Pietronero 1996). Whether there is a crossover to
homogeneity or not, the power law nature of the two-point and higher-order
correlation functions is itself suggestive of some kind of scaling behaviour
at least in some range. The simplest structure that obeys such a scaling law
is a single fractal.

A theoretical model describing a non-analytic inhomogeneous scale-invariant
universe is non-existent. The most-extensively-studied inhomogeneous
cosmological model is Tolman spacetime (Tolman 1934, Bondi 1947). Tolman's
dust solution has been used to model a hierarchical cosmology compatible
with the observational analysis of the redshift surveys (Bonnor 1974,
Ribeiro (1992a, 1992b), 1993). Recently, the Einstein equation for a
scale-invariant spherically symmetric inhomogeneous, but isotropic,
universe, which allows a non-vanishing pressure has been solved (Abdalla \&
Mohayaee 1997). However, the results obtained in this way are perturbative,
assume a preferred center for the universe and violate the linearity of the
Hubble law. A self-similar universe avoiding such difficulties can only be
constructed for a non-analytic distribution of matter. It is rather a
difficult task to construct a fractal metric and to solve Einstein equation
for a self-similar universe. However, many cosmological phenomena can be
accurately described by the Newtonian gravity, especially in the present
matter-dominated era.

In this work, we construct a self-similar universe whose dynamics is
governed by the Newtonian gravity. We divide the universe into $k$ spherical
clusters each of which contains $k$ subclusters which in their 
turn contain $k$ sub-subclusters. This clustering cascades 
down all the way to the level
of the galaxies which are at the lowest rung on the clustering ladder. The
mass and radius of each cluster can be used to define the fractal dimension
of our model.

We write the kinetic and potential energies of each cluster in terms of its
center of mass energy and the internal energies of its subclusters. The
thermal energy is obtained by requiring the total entropy of the canonical
ensemble of the clusters and their subclusters to remain constant. The final
Lagrangian is formulated in terms of two dynamical parameters: radius of the
largest cluster and its ratio to the radius of its
subclusters. The radius of the largest and smallest clusters, the ratio of
the mass to the critical mass contained in a sphere of radius $20$ Mpc, the
number of subclusters in each cluster and the ratio that characterises the
relative significance of thermal and gravitational energies are left as free
parameters. By fixing these to different observational values, we are
able to solve the equations of motion numerically using a Pascal program.
From the solutions, we can trace the evolution of the fractal dimension and 
verify the linearity of the velocity-distance relationship over time scales 
comparable to the age of the universe. 

The results are remarkable. We observe that for different initial conditions
a nearly homogeneous universe with a fractal dimension close to 3 evolves
into a universe with a fractal dimension of the order of 2 at the present
time. This fractal dimension fluctuates slightly about the value of 2 over
the future times but remains on the average constant. We also show that for
insignificant thermal energies, the Hubble law is closely obeyed by our
model.

We also make an estimate for the age of our self-similar universe. This is
one of the most challenging problems of Friedmann cosmology since the
observed age of the old stars in the globular clusters is far bigger than the
value estimated for the age of the universe in the nearly flat standard
model. The age of the universe obtained in our model is related to the
radius at which the crossover to homogeneity occurs. The farther the
crossover radius the older is the universe.

This article is organized as follows. In Section 2, we formulate our
clustering model. In Section 3, we obtain the kinetic, the potential and
the thermal energies, write the Lagrangian and the equations of motion. In
Section 4, we solve these equations numerically and discuss the validity and
limitations of our Newtonian approximation. In Section 5, we show different
plots of the fractal dimension for different choices of the initial
condition. Hubble law is discussed in Section 6. In Section 7, we study the
evolution of the scale factor and obtain a value for the age of the universe
in our model. Section 8 is devoted to the conclusion.

\section{A single-fractal model} 

\indent

We consider a hierarchical model of the universe, such that a spherical
cluster of radius $R_n$ and mass $M_n$ is composed of $k$ uniformly
distributed subclusters of uniform radius $R_{n-1}$ and mass $M_{n-1}$. That
is 
\begin{equation}
\label{fractal-definition}M_{n}=k M_{n-1}\quad {\rm and} \qquad R_{n}=\alpha
R_{n-1}\quad ,  
\end{equation}
where $\alpha$ stands for the ratio of the radius of the $n$th cluster to
the radius of the $(n-1)$th cluster. The parameters $k$ and $\alpha$ do not
depend on $n$.

The smallest constituent of the hierarchy is a galaxy of mass $M_1$ 
\footnote{Since galaxies do not expand, it is reasonable to 
assume that they do not contain subclusters.}. We suppose that the universe
contains $N$ hierarchies, where $N$ is a free parameter in this model (Fig.
1). 
\begin{figure}[htb]
\begin{center}
\leavevmode
\epsfxsize=4truecm{\epsfbox{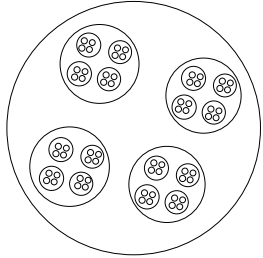}}
\vskip .5cm
\caption{{\it A simple example is shown 
of a hierarchical model in which $k=4$, $N=4$ 
and $\alpha\approx 3$.
}}
\end{center}
\end{figure} 

In terms of the mass M and radius R of the largest cluster, the expressions (%
\ref{fractal-definition}) can be rewritten as 
\be
R_n=\alpha^{n-N}R\quad ,\qquad M_n=k^{n-N}M\quad . \label{rr} 
\ee
The fractal dimension $D$ is given by (Coleman \& Pietronero 1992) 
\be
D=\frac{\ln k}{\ln \alpha}. \label{fractaldim}  
\ee

We assume that the clusters neither dissociate nor collide during the
evolution which leaves $R$, $\alpha$ and consequently the fractal dimension
as the only dynamical quantities of this model.

\section{The Lagrangian} 

\indent

In this section, we write down the Lagrangian of our model by computing the
gravitational and thermal energies of the clusters.

The kinetic energy of a sphere of radius $L$ expanding uniformly is 
\be
T = \frac{1}{2}\int_{0}^{L} \rho \dot{r}^2 dv= \frac 3{10} M\dot L^2\quad, 
\label{kinetic-energy} 
\ee
where we have used 
\footnote{Note that the Hubble law is not assumed by
this expression since $\dot L/ L$ is allowed to be 
different for different clusters. } 
\be
\dot r={\frac{\dot L}{L}} r. 
\ee
For a composite system, such as ours, the total kinetic energy is the sum of
the center of mass kinetic energy and the kinetic energies of the
constituents of the system in the center of mass frame. Therefore, the
kinetic energy of the $(n+1)$th cluster in our model can be written as 
\begin{equation}
T_{n+1}=\frac{3 M_{n+1} \dot R_{n+1}^2}{10}+k T_n, 
\end{equation}
Subsequently the total kinetic energy $T$ is 
\begin{equation}
\label{kinetic}T=\frac{3}{10}\sum_{n=2}^{N} k^{N-n}M_{n}\dot{R_{n}}^2
\approx \frac{3}{10}MR^2(\frac{\dot{R}^2}{R^2}-\frac{2 \dot{\alpha} \dot{R} 
}{\alpha^3 R}+\frac{\dot{\alpha}^2}{\alpha^4}) 
\end{equation}
where the approximation is made for $\alpha^2$ much bigger than unity 
\footnote{
This approximation is well within the range given 
by the observational data. For 
example, the radius of the local supercluster,
is about 10 times the radius of the local group.}.

The potential energy can be derived in a similar manner. The gravitational
potential energy of a uniform sphere is $-{3GM^2}/{5R}$ which becomes $-{%
3GM^2}(1-1/k)/5R$ for a discrete mass distribution. In our case the
potential energy of the $(n+1)$th cluster is 
\begin{equation}
\label{potential-sn+1}U_{n+1}=kU_{n}-\frac{3 G M_n^2 k(k-1)}{5 R_{n+1}}\quad
.  
\end{equation}
We also note that 
\begin{equation}
\left\{ 
\begin{array}{ll}
D>1 & \mbox{for $(\frac{\alpha}{k})^{N-1} \ll 1$} \\  &  \\ 
D<1 & \mbox{for $(\frac{\alpha}{k})^{N-1} \gg 1$} 
\end{array}
\right. 
\end{equation}
The first range is compatible with the values of the fractal dimension given
by the observations 
\footnote{The 
fractal dimension lies between 
$1.2$ (Peebles 1980) and $2.0$ (Pietronero et al. 1997).} and can be used to
approximate the potential energy to, 
\begin{equation}
\label{potential}U\approx -\frac{3GM^2}{5\mu R(1-\alpha/k)}, 
\end{equation}
where $\mu= (1-k^{-1})^{-1}$.

As well as expanding, the clusters are also fluctuating randomly. This
introduces a generalized thermal term into the Lagrangian. The canonical
partition function for a system of ${\cal N}$ particles interacting through
gravitational potential is given by 
\begin{equation}
{\cal Z}=\frac{1}{{\cal N}!}\int dq^{3{\cal N}}dp^{3{\cal N}}
\exp\left(-\beta {\cal H}(p,q)\right), 
\end{equation}
where the Hamiltonian ${\cal {H}}$ is \be
{\cal H}=\sum_{i=1}^{{\cal N}}\left({\frac{P_i^2}{2m}}\right)+U\{q_i\} \ee 

In order to compute the integral over the potential term we observe that the
volume available to one subcluster inside the $(n+1)$th cluster is $\frac{4}{%
3}\pi(R^3_{n+1}-(k-1)R^3_n)$. For $k$ subclusters, there is an exponent of $k
$ for the complete available volume in the configuration space, thus 
\begin{equation}
{\cal {Z}} \propto \beta^{-\frac{3{\cal N}}{2}} \prod_{n=2}^{N}
(R^3_n-(k-1)R^3_{n-1})^{(k^{N-n+1})} e^{-\beta U}. 
\end{equation}
To obtain the above expression we have factored out $e^{-\beta U}$ as a
constant, where we have assumed that its dependence on $q$ has already been
accounted for by taking a homogeneous distribution of subclusters. Using the
standard equations of thermodynamics for the conserved entropy $S$: 
\begin{equation}
S=\ln{\cal {Z}}- \beta \frac{\partial \ln{\cal {Z}}}{\partial \beta} 
\end{equation}
the thermal energy can be written as 
\begin{equation}
\label{thermal}{\cal {E}}\approx {\frac{3M a}{10}} R^{-2\mu} (\alpha^3 -k
+1)^{-2\mu /3} \alpha^{3p}  
\end{equation}
for $k^{N-1}\gg 1$, implied by $\alpha^2 \gg 1$, $p=\frac{2}{3}(N\mu +\mu^2)$
and the value of the constant $a$ depends on the initial value of the
large-scale entropy.

Putting expressions (\ref{kinetic}), (\ref{potential}) and (\ref{thermal})
together, the Lagrangian, is given by \be
{\cal {L}}={\frac{3M}{10}} \left(\dot R^2-\frac{2\dot \alpha \dot RR}{\alpha
^3}+\frac{\dot \alpha^2R^2}{\alpha ^4}+ \frac g{(k-\alpha )R}-aR^{-2\mu
}(\alpha ^3-k+1)^{-2\mu/3} \alpha ^{3p}\right), \label{lagran} \ee
where $g={2GMk}/\mu$. The equation of motion for $\alpha$ becomes 
\begin{eqnarray}
\ddot \alpha &=& 2\frac{\dot \alpha ^2}\alpha +\alpha (\frac{\dot R}R)^2- 2
\frac{\dot \alpha \dot R}{R}+\frac{\alpha g(\alpha ^3-k)}
{2(k-\alpha )^2R^3}\nonumber\\
&+& 
\left(aR^{-2\mu -2}\alpha^{3p+3}(\alpha ^3-k+1)^{-\frac{2\mu}{3}-1}\right)
\left(\mu \alpha^3 -\frac{3}{2} p (\alpha^3-k+1)\right)
\end{eqnarray}
while for $R$ we obtain 
\begin{eqnarray}
\ddot R&=&\frac{\alpha^{-2}\dot{R}^2}{R}-2\alpha^{-3}\dot{\alpha}\dot{R}+
\frac{g(2\alpha -k)}{2(k-\alpha )^2R^2}\nonumber\\
&+& 
\left(aR^{-2\mu -1} 
(\alpha^3-k+1)^{-\frac{2\mu}{3}-1}\alpha ^{3p} \right)
\left(
\mu \alpha^3 +(\mu -\frac{3p}{2})(\alpha^3-k+1)\right).
\end{eqnarray}

\section{Numerical solutions and their limitations} 

\indent

In this section, we solve the equations of motion numerically. The
parameters in our model are $R, \dot R, \alpha, \dot\alpha, N, k, a$ and $M$%
. The first four parameters are taken as initial conditions 
\footnote{We take the present time to be the zero time and evolve our 
model back and forth in the matter-dominated era.} and the latter four as
constants. These are all expressed in terms of observational quantities.

The crossover radius $R$ marks the transition from a hierarchical
distribution to a random homogeneous distribution and we take it to lie in
the range 30-1000 Mpc. Subsequently, the present value of $\dot R$ is fixed
by the observed value of the Hubble constant which is about $55\,\, {\rm km}%
\,\,{\rm sec}^{-1} {\rm Mpc}^{-1}$ (Tammann 1997). Furthermore, the 
vanishing of 
$\dot\alpha$ at near distances is implied by the Hubble law. However, 
whether this law remains valid
or not over large distances is not determined by this assumption 
and shall be shown in Section 6.

The parameters $\alpha, N, k$ and $M$ can be fixed by using the radius of
the largest and smallest clusters ($R$ and $R_2$, respectively, where the
latter is taken to be about $3$ Mpc which is the size of our local group 
\footnote{Our results basically do not change for 
larger clusters such as 
Virgo cluster which has a radius of about 10Mpc.}), $\Omega$ which is
defined to be the ratio of the mass contained in a sphere of radius 20 Mpc
to the critical mass and is taken to lie in the range 0.25--1.0 (Ostriker
1993), the typical mass of a galaxy which is about $10^{11}$ times the solar
mass and the present value of the fractal dimension which is about 2
(Pietronero et al. 1997).

The parameter $a$ can be fixed by using the hierarchical level where the
virialization sets up, which is to say when $2{\cal E}_{N_c}=-U_{N_c}$. The
virialization level $N_c$ marks the relative weight of the gravitational
potential energy and the thermal kinetic energy and replaces the parameter $a
$ in our computations. The center of mass thermal energy of each cluster is $%
3/2\beta= {\cal E}/{\cal N} $. Hence, the center of mass thermal energy of
subclusters contained in a typical cluster turns out to be ${\cal E}%
_n=k^{2-N}{\cal E}$ . Thus, the thermal energy tends to zero
when the parameter $N_c$ approaches minus infinity.

It is also worth commenting on the Newtonian and various other
approximations that we have used in our model. In general the Newtonian
approximation can be satisfactorily used in the matter-dominated era.
However, in a self-similar model 
the range of such an approximation can be further constrained. That 
is to say as long as
the time needed for light to cross the system is small compared to the
characteristic time of the evolution of the dimension the Newtonian regime
is applicable. Specifically, for times later than about $-15\times 10^9\,\, 
\mbox{years}$, or equivalently for $\frac{c\dot{\alpha}}{R\alpha} \ll 1$, 
the use of the non-relativistic approximation is justified. Another 
approximation has been the neglect of $\alpha^{-2}$, which
results in a maximum error of about $9\%$ in our results for $D\approx 3$.

\section{Evolution of the fractal dimension and CMBR isotropy} 

\indent

In this section, we study the evolution of the fractal dimension by using
our numerical solutions for $\alpha $, obtained in the last section. The
graphs of fractal dimension versus time given at the end of this article
show the evolution of the fractal dimension, back in the matter-dominated
era, until the time when the Newtonian approximation fails. 
All the
graphs are plotted for three initial 
dimensions of $D=1.5,\,2.0$ and $2.5$.

Comparing different graphs in Figure 3 we see that for shorter crossover
radii the initial fluctuations die away faster and the fractal dimension 
settles at its almost constant value of 2 earlier.

The homogeneity at earlier times, due to the significantly large thermal
energy, is remarkable since it conforms with the isotropy of the microwave
background radiation 
\footnote{We can only extrapolate our model back
to the time of decoupling since the
Newtonian approximation cannot be 
used beyond approximately $-15$ Gyr.}.

We also see that the transition from homogeneity to fractality is marked by
oscillations. The frequency of these oscillations increases with increasing
initial dimension and thermal energy and decreases with time because of the
cooling caused by the adiabatic expansion of the universe. The reason why
this oscillatory behaviour does not appear for the two lower initial 
dimensions is
the breakdown of the Newtonian regime which occurs earlier for smaller 
initial dimensions (see Section 4).

A qualitative description of these oscillations can be made by studying the
potential terms of the Lagrangian (\ref{lagran}). The thermal term in the
potential energy consists of a decreasing function of the parameter $\alpha$%
, which diverges as we approach homogeneity, and a fast increasing function
of $\alpha$. Therefore, a minimum exists near $D=3$ which is the origin of
the observed oscillations. The gravitational potential term becomes
important when $D$ approaches unity. The negative sign of this term smoothes
down the sharp increase of the potential near $D=1$ which accounts for the
decrease of the frequency for small values of the initial dimension 
({\it i.e.}, $D=1.5$ and $D=2.0$).

The solutions are quite stable for future times and the dimension remains at
a mean value of about 2 over sufficiently long times and for a complete
range of parameters. This confirms that at the present time the dimension is
almost time-independent which was already assumed in solving the equations
of motion.

\section{The Hubble law} 

\indent

In the previous section, we have discussed the evolution of the fractal
dimension and the isotropy of the microwave background radiation. 
This one is devoted to Hubble law.

In the preceding sections, we have assumed Hubble law to be valid at small 
distances. The uniform distribution of subclusters contained in each cluster
implies the Hubble law for that cluster. However, this does not imply the 
Hubble law for a full range of distances. Using expressions (\ref{rr}) and 
(\ref{fractaldim}) we obtain 
\begin{equation}
\dot R_n=R_n \left(\frac{\dot R}R-\frac{\dot D}D \ln 
\left(\frac{R_n}R\right)\right) 
\end{equation}
for the velocity-distance relationship. Although the relationship is valid
only for a discrete set of $R_n$'s, one can interpolate the velocity for the
intermediate values since the observed velocity distribution is not
expected to be discontinuous. In addition, because of the finiteness of the
speed of light the value for $\dot D/D$ is calculated in the retarded time
which is $t-R_n/c$ for a typical galaxy located approximately at the center
of a cluster 
\footnote{In fact,  $\dot D/D$ is only calculated for galaxies which are 
located approximately about the center of a cluster.
For galaxies located near the edge of the clusters the retarded time is 
direction dependent and does not have a single value.}.

Comparing graphs in Figure 4, we see that decreasing crossover radius leads
to a better fit for the Hubble law, which is compatible with the results of
standard cosmology.

The presence of the second term in the last equation causes the $\dot R_n$
to vanish for large $|\dot D/D|$. Increasing the thermal energy always
results in a vanishing $\dot R_n$ at a certain distance which increases with
lowering $\Omega $ or increasing $R$ (Fig. 4). The fact that the Hubble law
is so well-fitted up to some hundred megaparsecs clearly rules out the
earlier measurements of the dimension which was about $1.2$ (Peebles 1980).
In all of the graphs except for the case of $D=2.0$ a linear behaviour is
obtained which also confirms the recent observational results of $D=2.0\pm
0.2$ (Pietronero et al. 1997).

For relatively small thermal energies ({\it i.e.}, when $N_c=0$) the
linearity of the Hubble law is more robust although the graph for $D=1.5$
shows deviations from the Hubble law at large distances. In general, smaller 
$\Omega$ and $R$ give better fits for the Hubble law. This is because the
behaviour of the universe in our model is mainly determined by the kinetic
rather than the thermal term in the Lagrangian (\ref{lagran}). Furthermore,
in the absence of gravitational and thermal energies, the universe expands
self-similarly with a constant dimension which, ignoring the
time-retardation effect, leads to a perfect Hubble law.

\section{The age of the universe}  

\indent

In the preceding sections, we have discussed the isotropy of the microwave
background radiation and the Hubble law. The other crucial constraint, set
by observations, on all viable cosmological models, is the age of the
universe. The lower bound to the age of the universe, obtained from the age
of the oldest stars in the globular clusters, lies far above the value
estimated by the nearly flat standard model (Bolte \& Hogan 1995). 
In this section, we make an estimate for the age of the universe 
in our model.

Any estimate of the age of the universe is highly dependent on the value of
the parameter $\Omega$ which in our model depends on the crossover radius. A
higher value of $\Omega$ leads to a younger universe and vice versa. In our
model, $\Omega$ as defined in the standard model does not exist. However, we
have assumed that $\Omega$ lies in the range $0.1-1$ as given by the
Friedmann model which is only valid for a crossover radius of about $20$
Mpc. A larger crossover radius leads to a much lower value for $\Omega$ and
subsequently to a much older universe.

Taking the extreme case of $\Omega=1, N_c=0$ and 
a crossover radius of $300$
Mpc we obtain $15-17$ Gyr for the age of the universe 
\footnote{Extrapolation
of the first graph in Fig. (2) gives 
this value.}. This result is in agreement
with the maximum age estimated for the old 
stars which is $15.8\pm 2.1 $
Gyrs (Bolte \& Hogan 1995). On the other 
hand, if we take the lower value of
30 Mpc for the crossover radius then we 
obtain $13$ Gyrs for the age of the
universe which lies below the astrophysical 
lower bound (see Fig. 2). 
\vskip -1cm
\begin{figure}[ht]
\begin{center}
\leavevmode
\begin{eqnarray}
\epsfxsize= 7truecm{\epsfbox{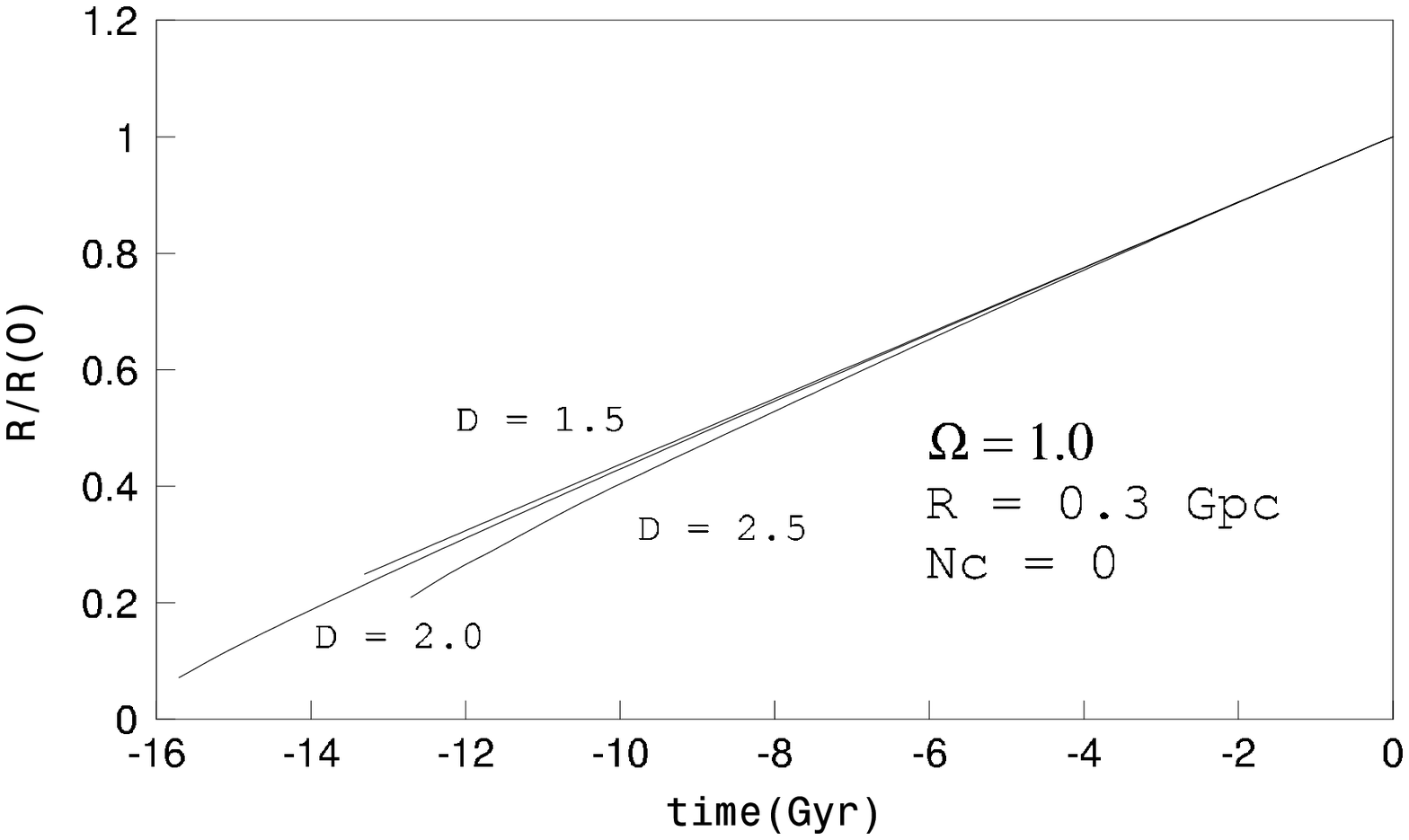}}\hskip -0cm & &
\epsfxsize= 7truecm{\epsfbox{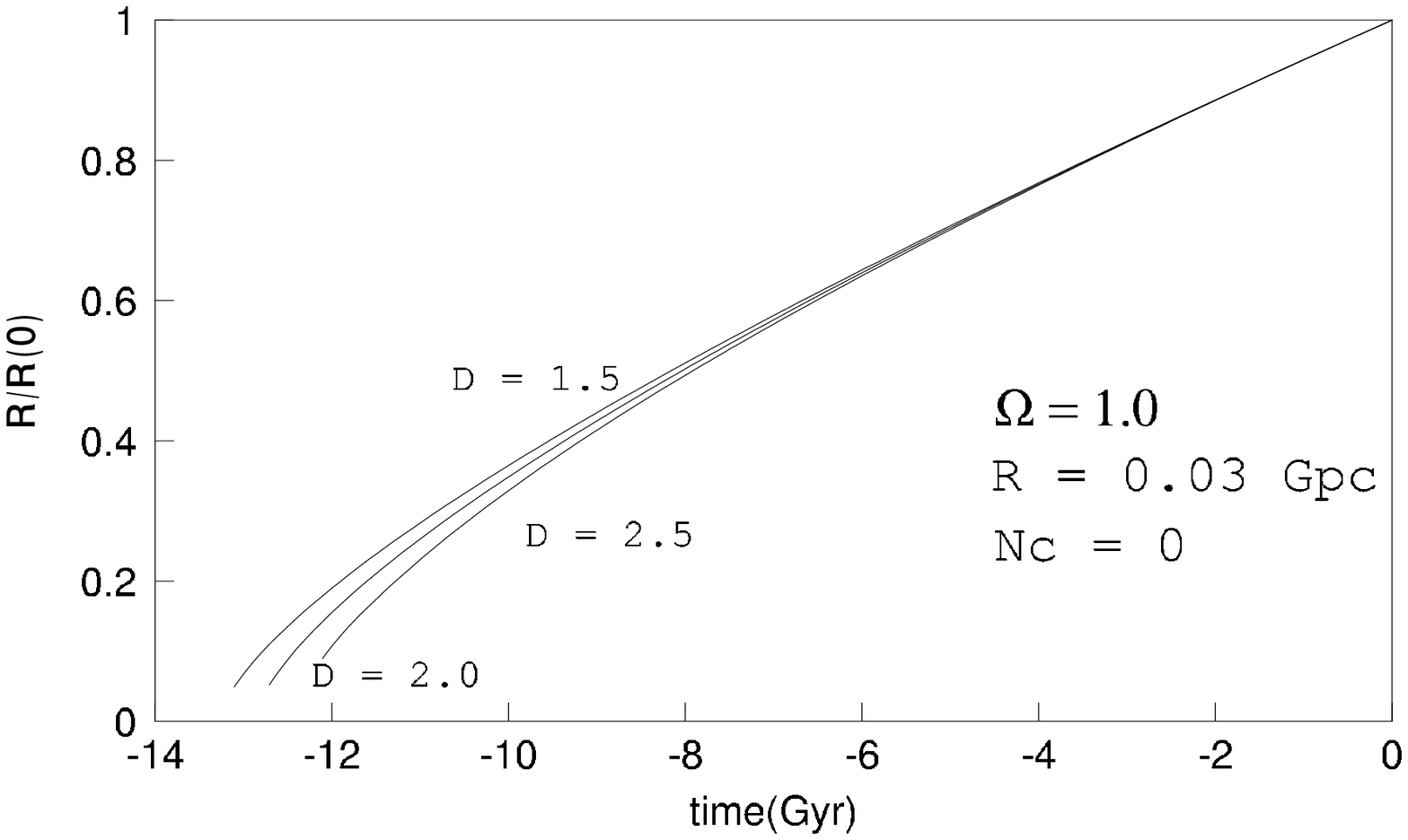}}\nonumber
\end{eqnarray}
\vskip -1cm
\caption{{\it
The radius of the universe is plotted against time for
different initial conditions.          
}}
\end{center}
\end{figure}             

\vskip -2cm
\section{Conclusion}


\indent

We have constructed a model for a nonrelativistic fractal universe. Our
model starts off homogeneous and evolves rapidly to a self-similar universe
with a remarkably constant fractal dimension of about 2. The homogeneity at
the earlier times explains the isotropy of the microwave background
radiation. We have also shown that the Hubble law is closely obeyed by our
model for small thermal energies. We have estimated the age of the universe
and have shown that it complies with the corresponding observational
results. It remains an open problem to extend our model to multi-fractals and
to the relativistic regime.

{\bf Acknowledgements} We thank R. Mansouri and M. Khorrami for useful
discussions. This work  has been partially supported
by Conselho Nacional de Desenvolvimento Cient\'\i fico e Tecnol\'ogico,
CNPq, Brazil, and Funda\c c\~ao de Amparo \`a Pesquisa do Estado de
S\~ao Paulo (FAPESP), S\~ao Paulo, Brazil.


\vfill\eject
\vskip -1cm 
\begin{figure}[ht]
\begin{center}
\leavevmode
\vskip -3cm
\begin{eqnarray}
\epsfxsize= 8truecm{\epsfbox{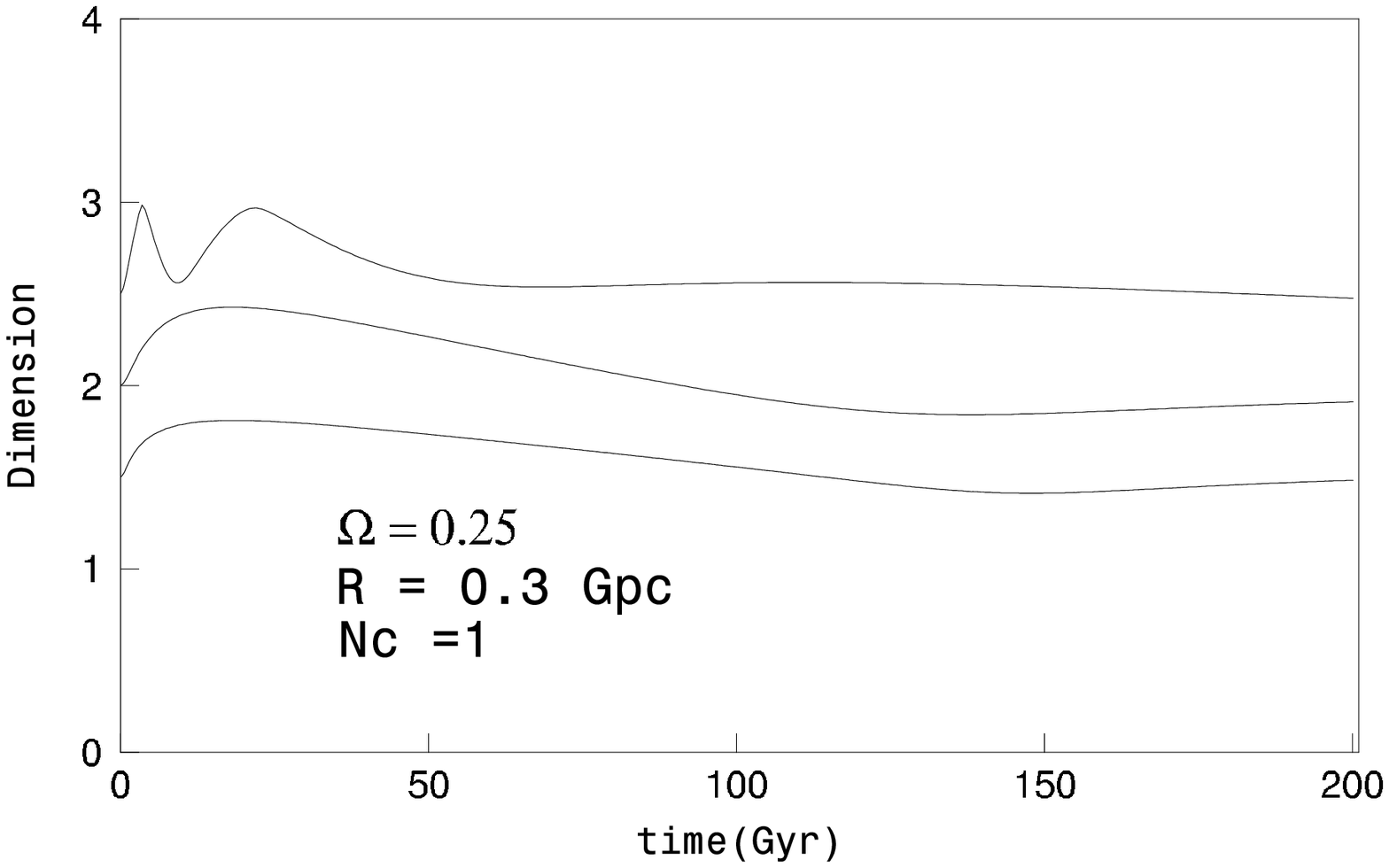}}\hskip  -0cm & & 
\epsfxsize=8truecm{\epsfbox{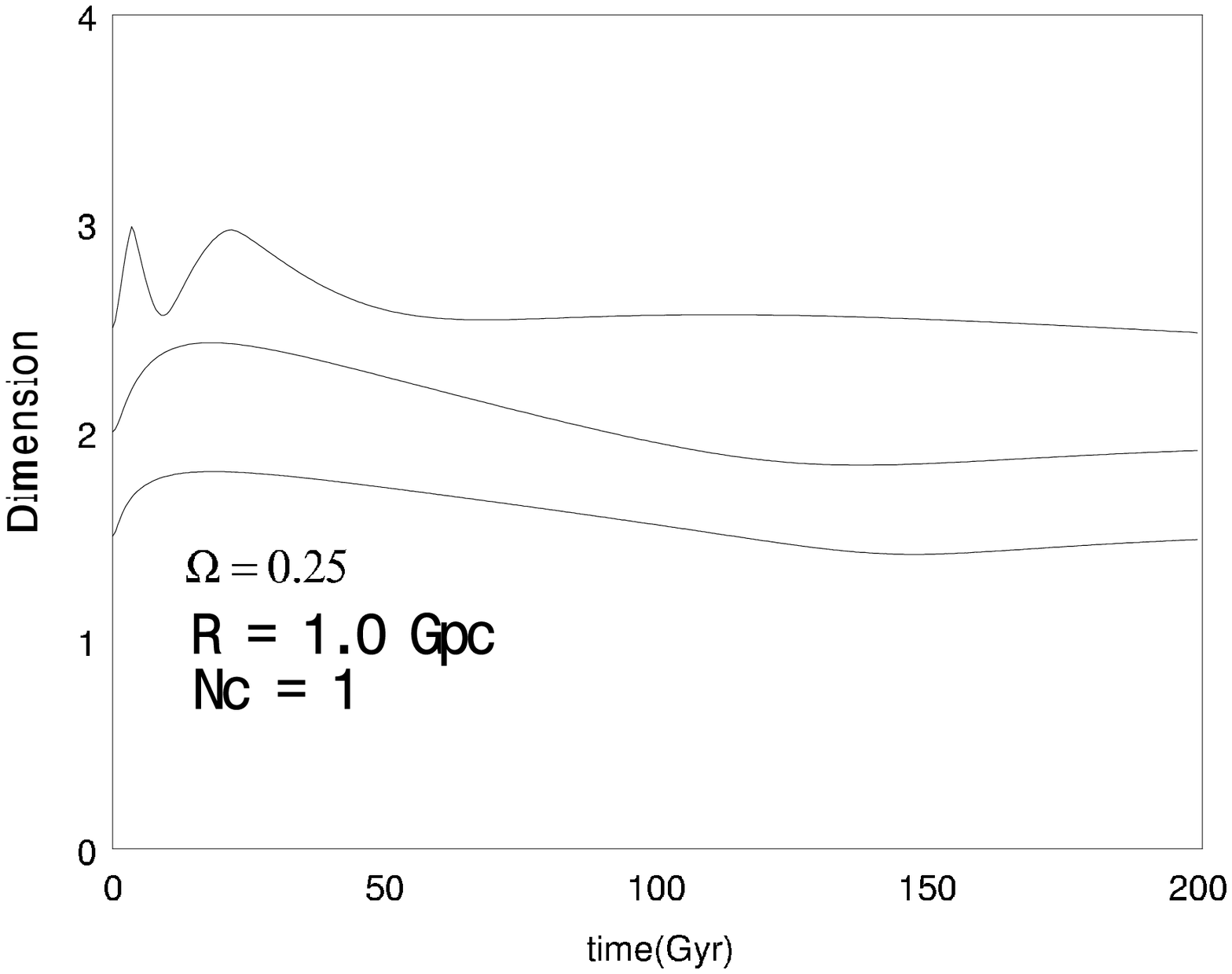}} \nonumber\\
\epsfxsize=8truecm{\epsfbox{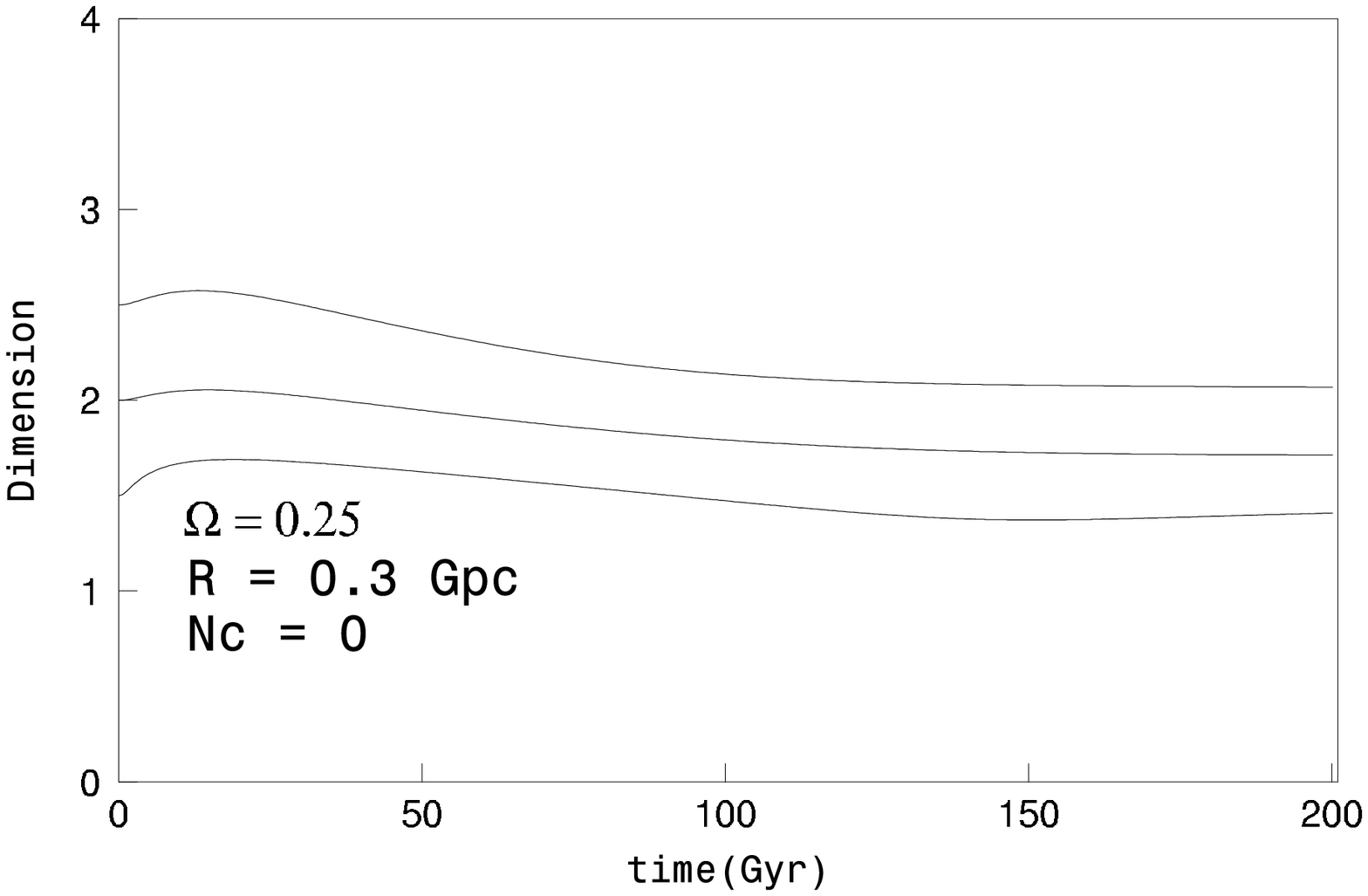}} \hskip -0cm & &
\epsfxsize=8truecm{\epsfbox{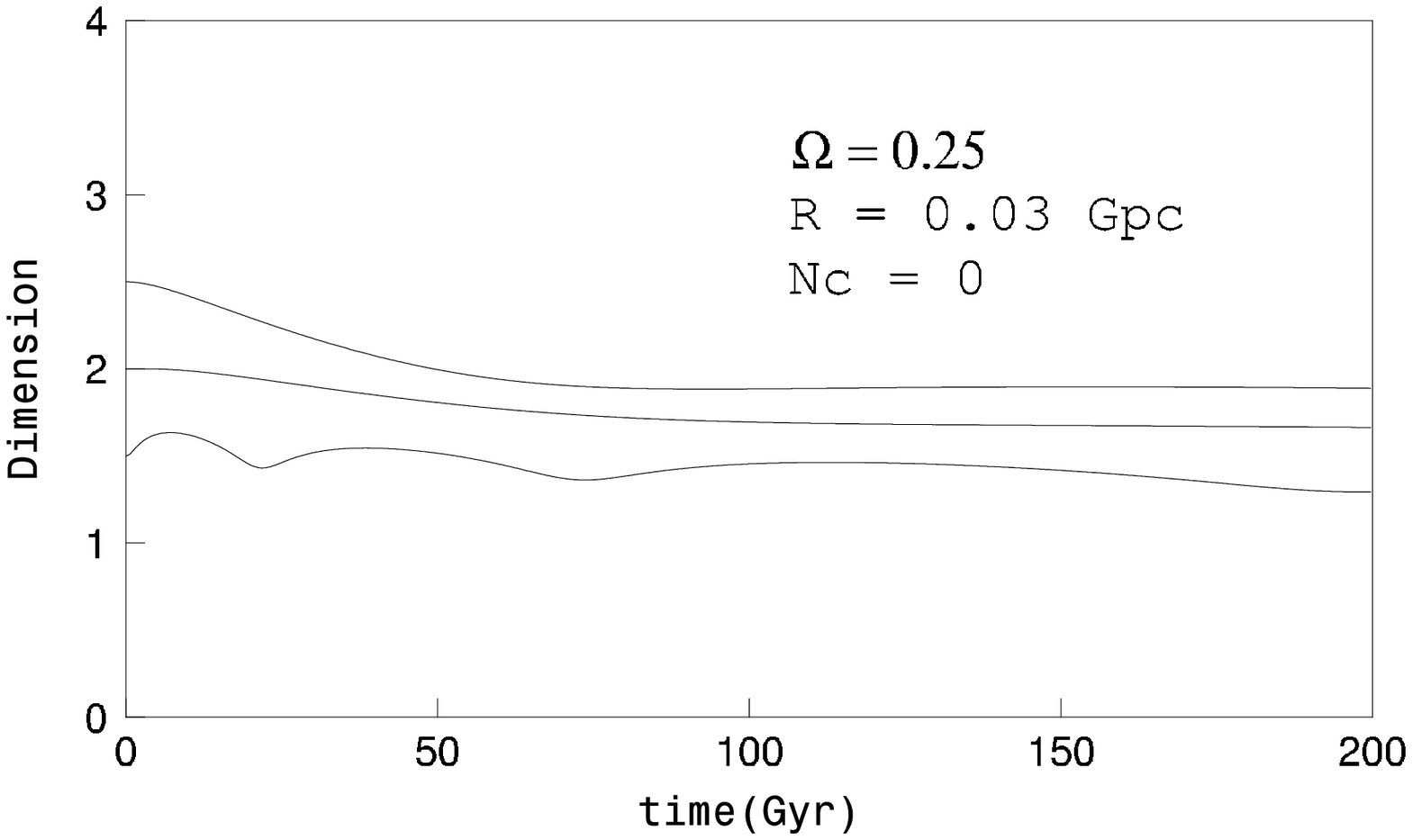}}\nonumber \\
\epsfxsize=8truecm{\epsfbox{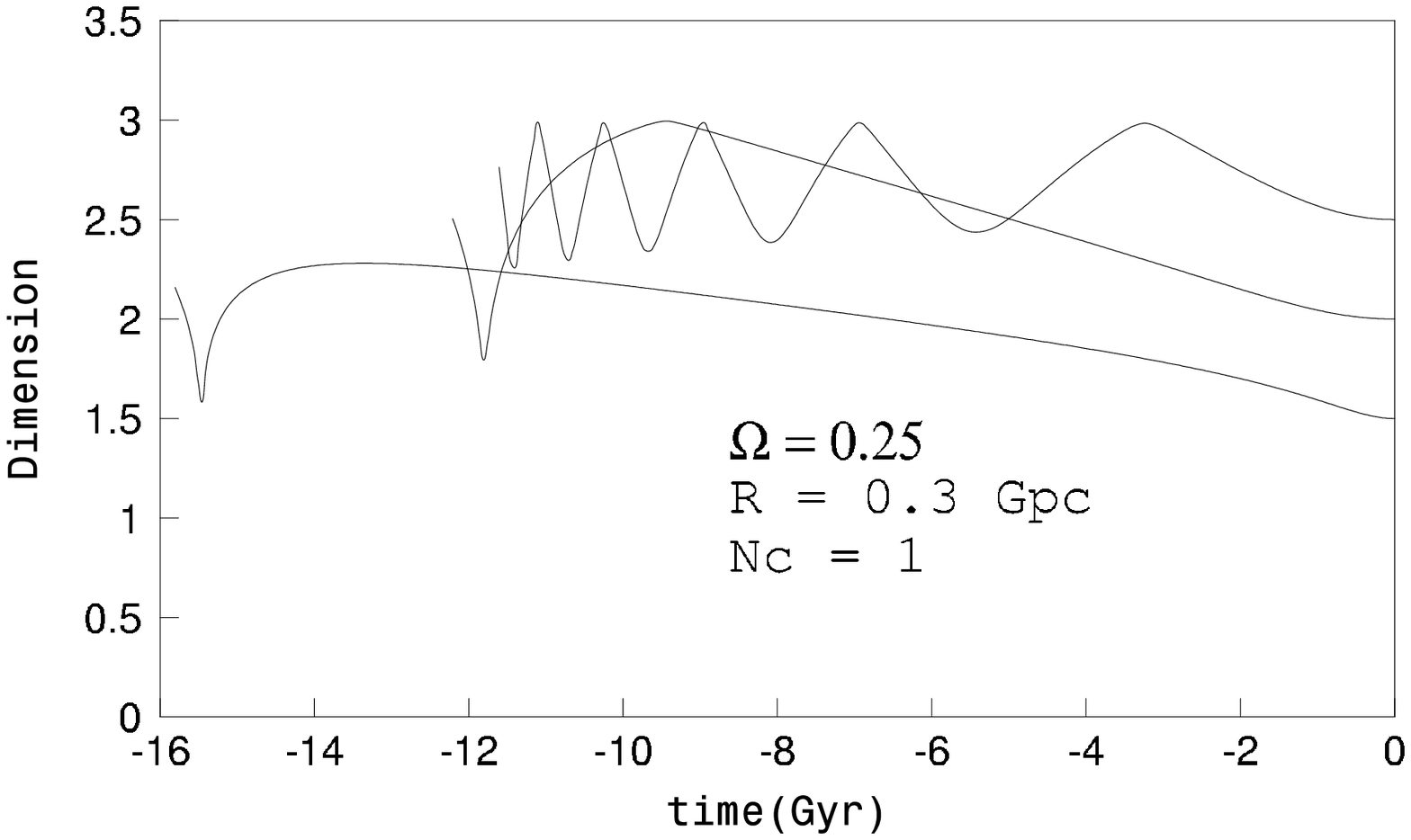}}\hskip -0cm & &
\epsfxsize=8truecm{\epsfbox{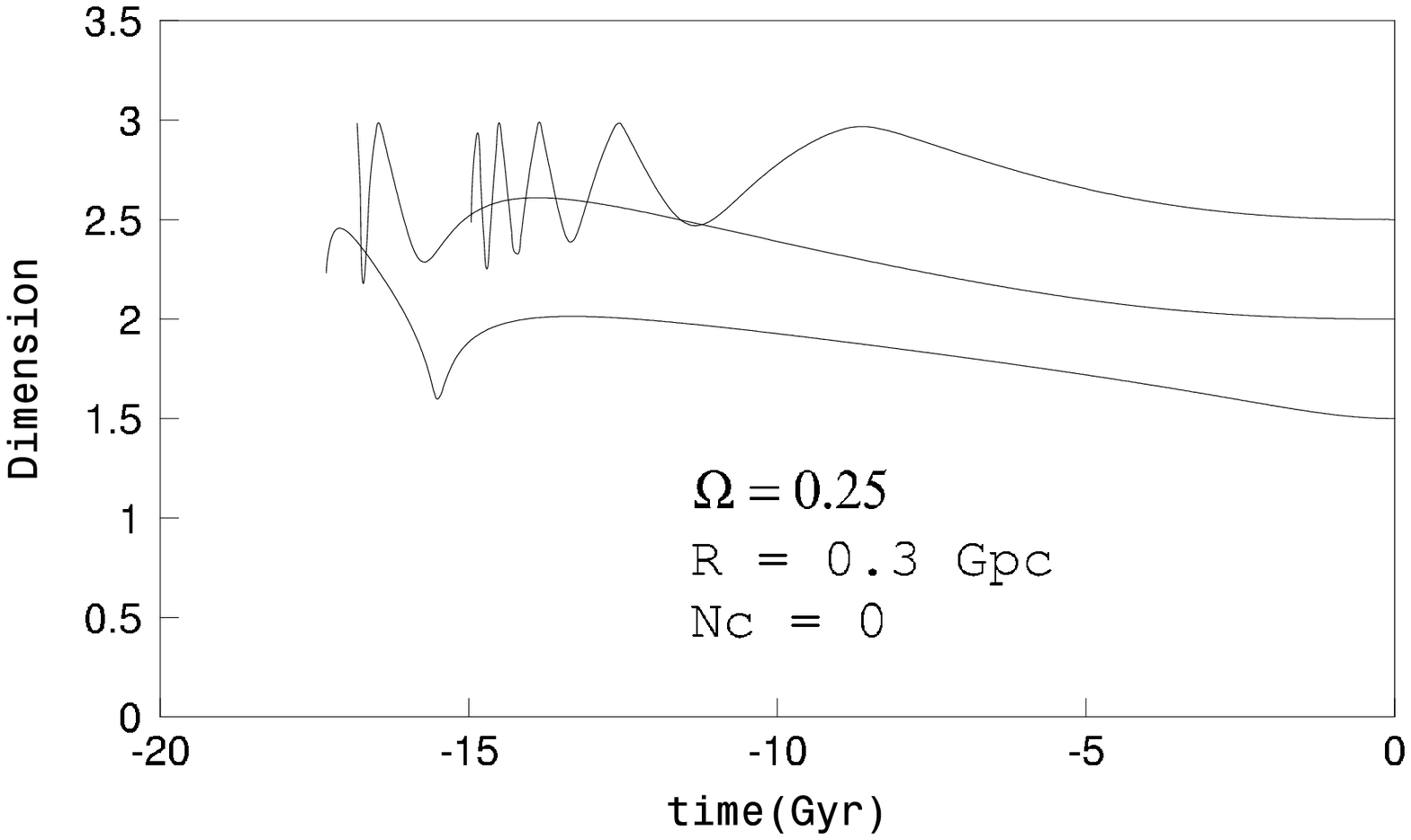}}\nonumber \\
\hskip 0 cm 
\epsfxsize=8truecm{\epsfbox{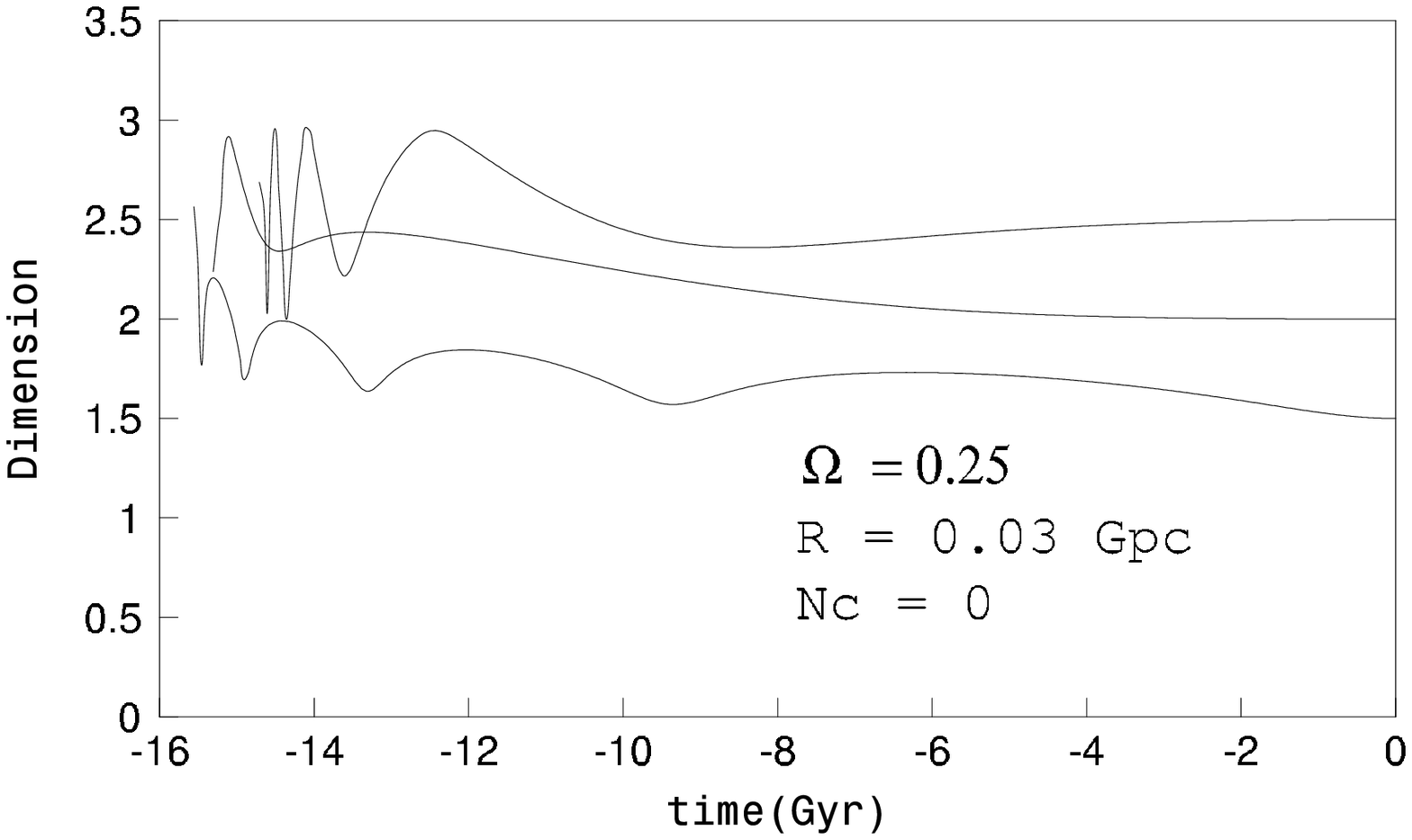}}\nonumber
\end{eqnarray}
\vskip -0cm
\caption
{{\it
Evolution of the fractal dimension for different initial dimensions,
crossover radii $R$, virialization levels $N_c$ and parameters $\Omega$.
}}
\end{center}
\end{figure}
\vfill\eject
\vskip -1cm 
\begin{figure}[ht]
\begin{center}
\vskip -3cm
\leavevmode
$
\begin{array}{lll}
\epsfxsize= 5truecm{\epsfbox{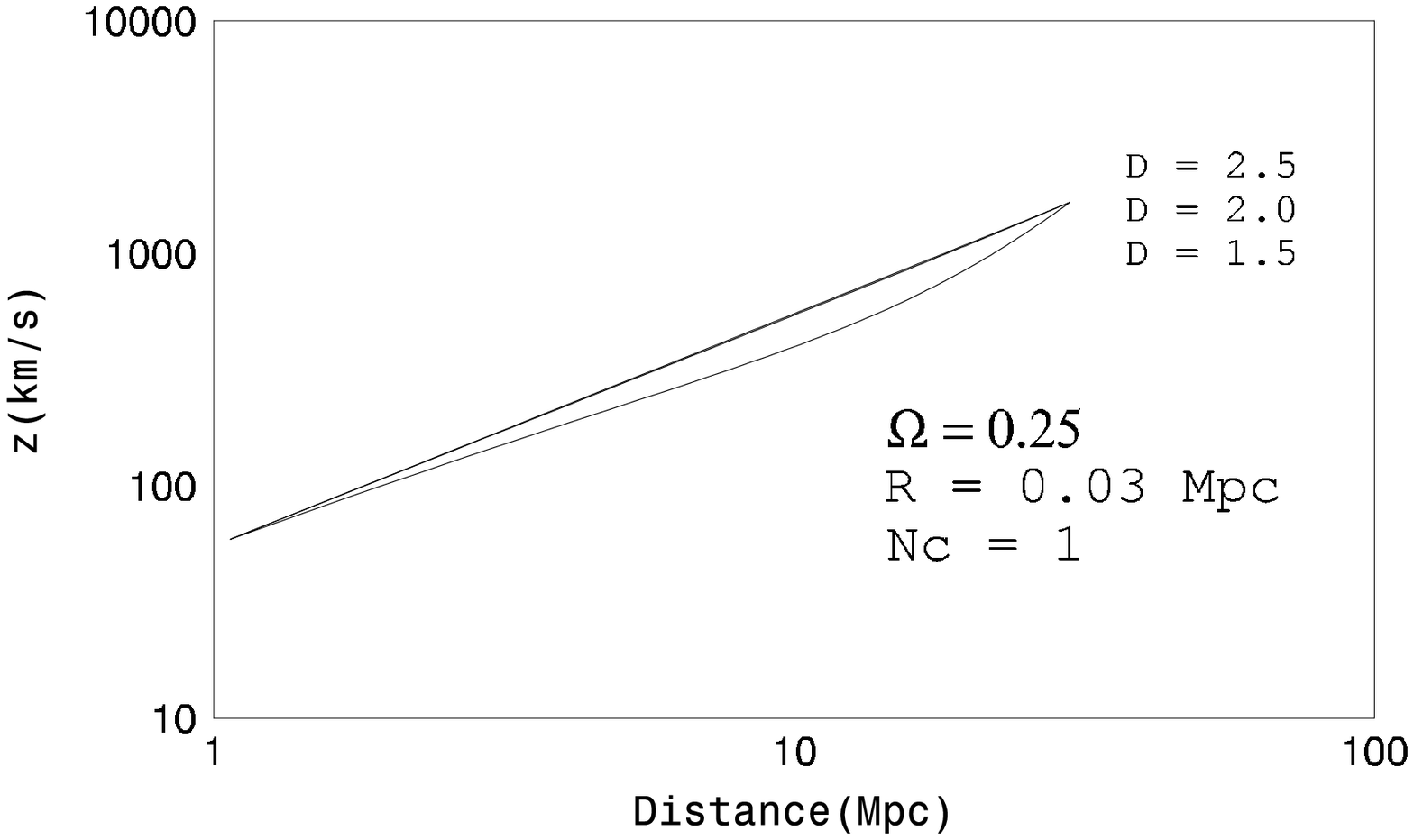}}& 
\epsfxsize= 5truecm{\epsfbox{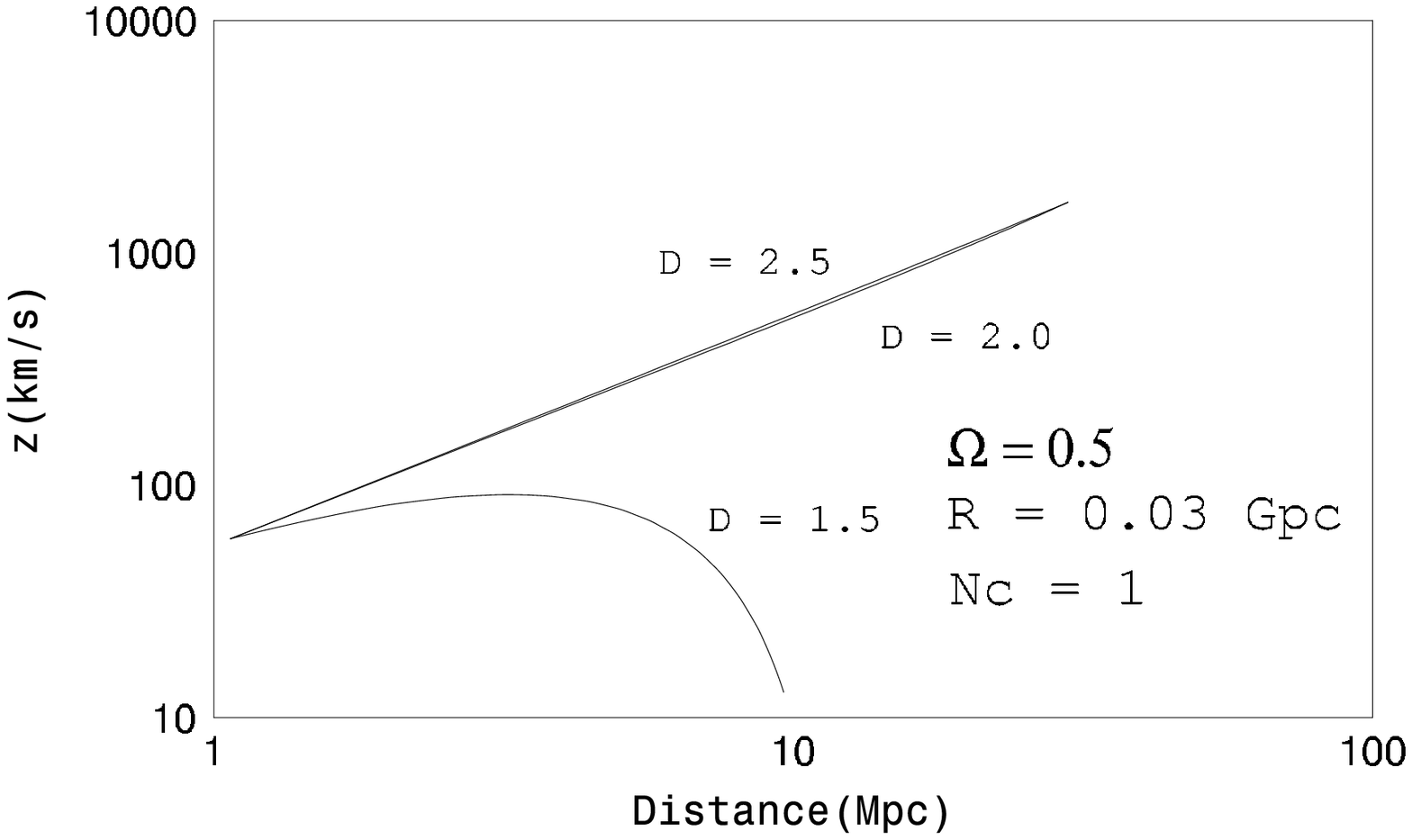}}  & 
\epsfxsize= 5truecm{\epsfbox{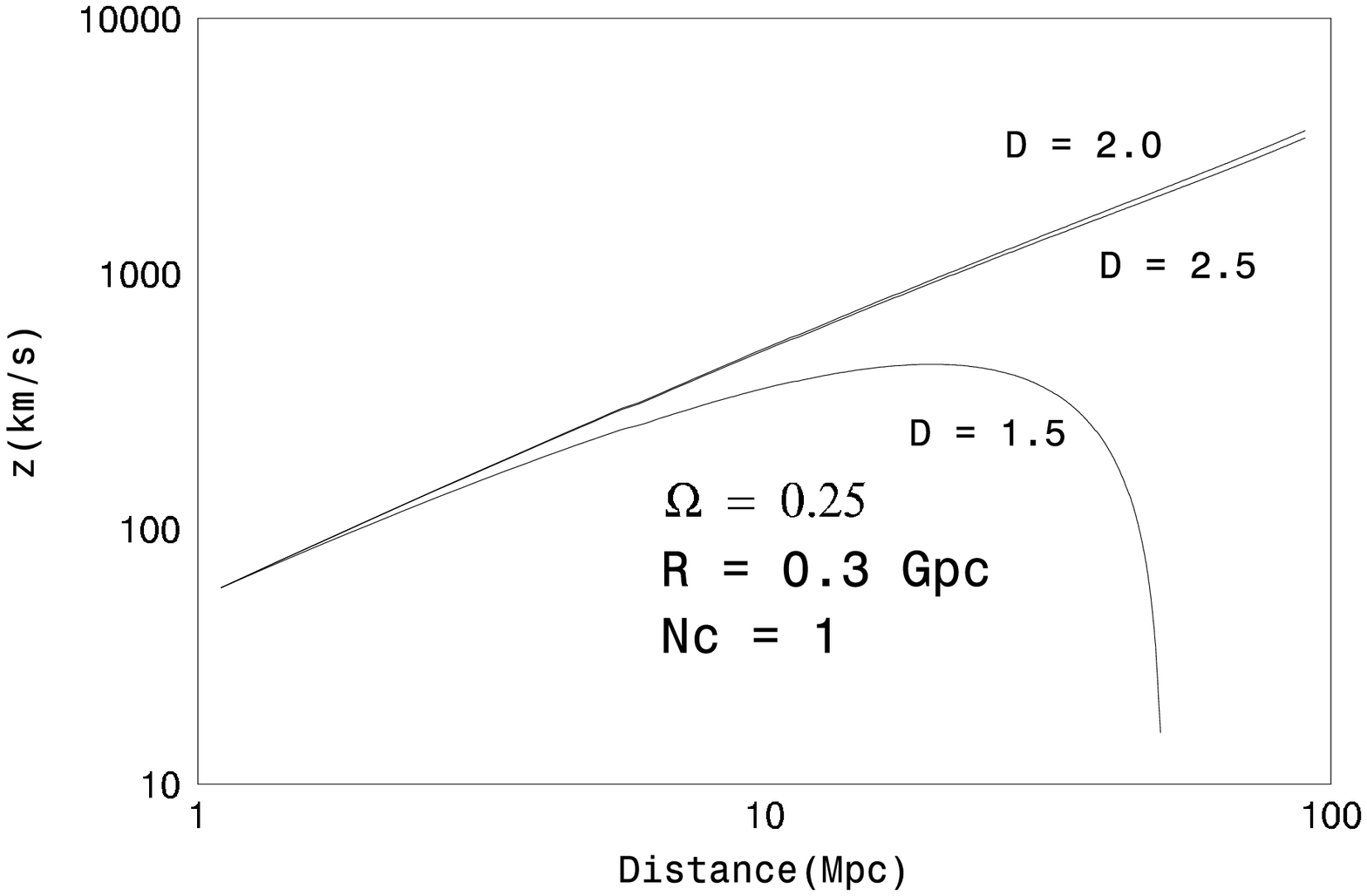}} \\
\epsfxsize= 5truecm{\epsfbox{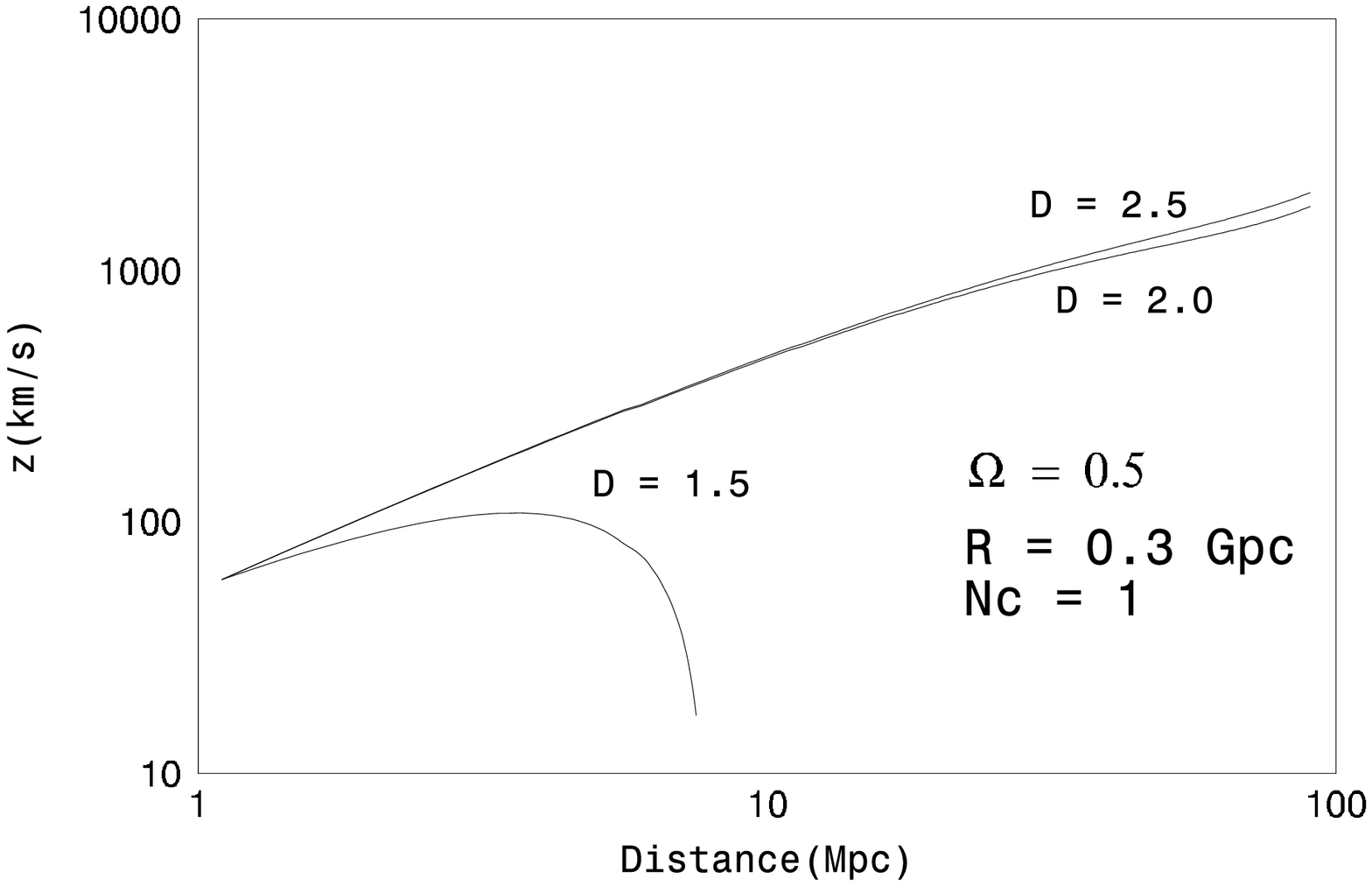}}  & 
\epsfxsize= 5truecm{\epsfbox{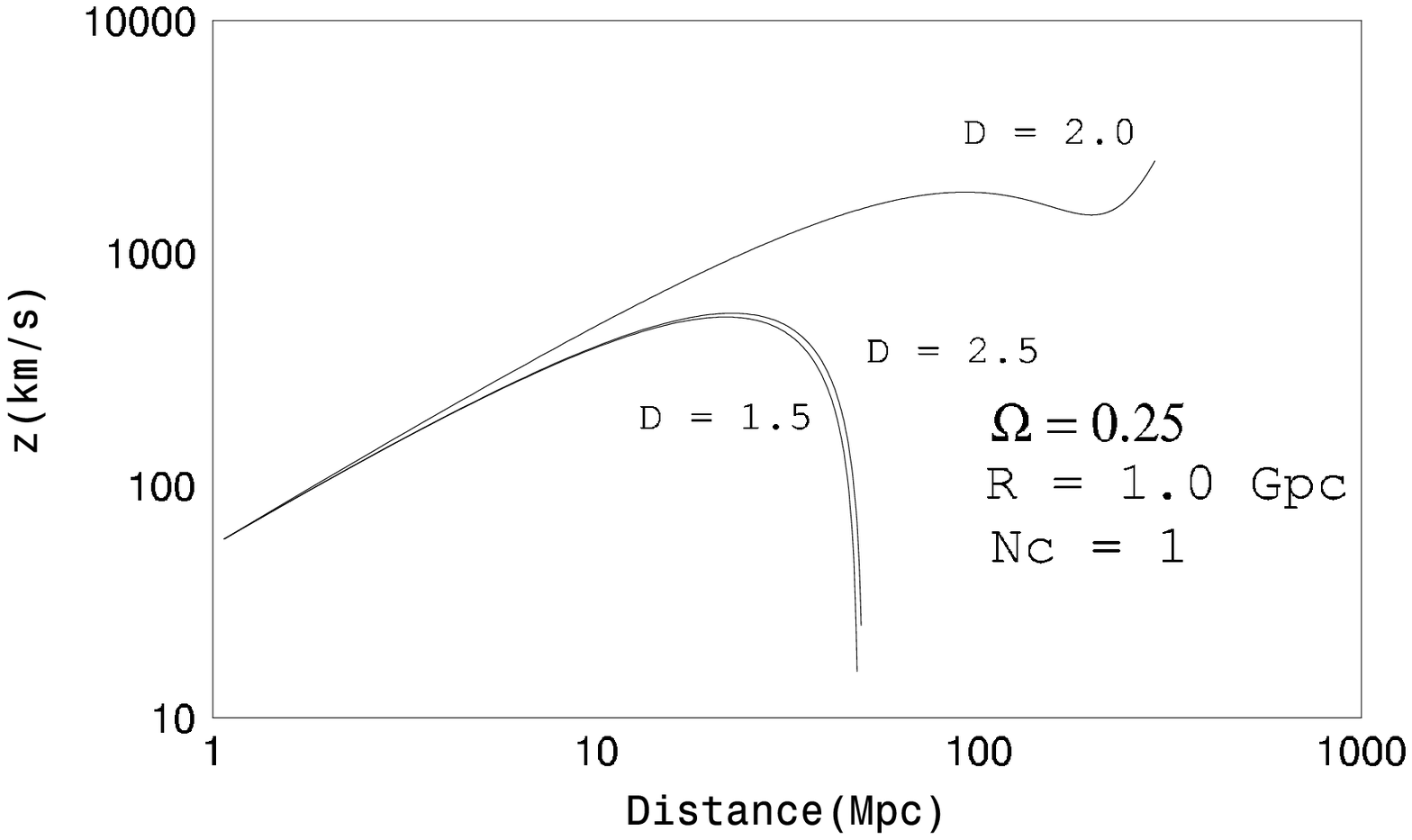}}  & 
\epsfxsize= 5truecm{\epsfbox{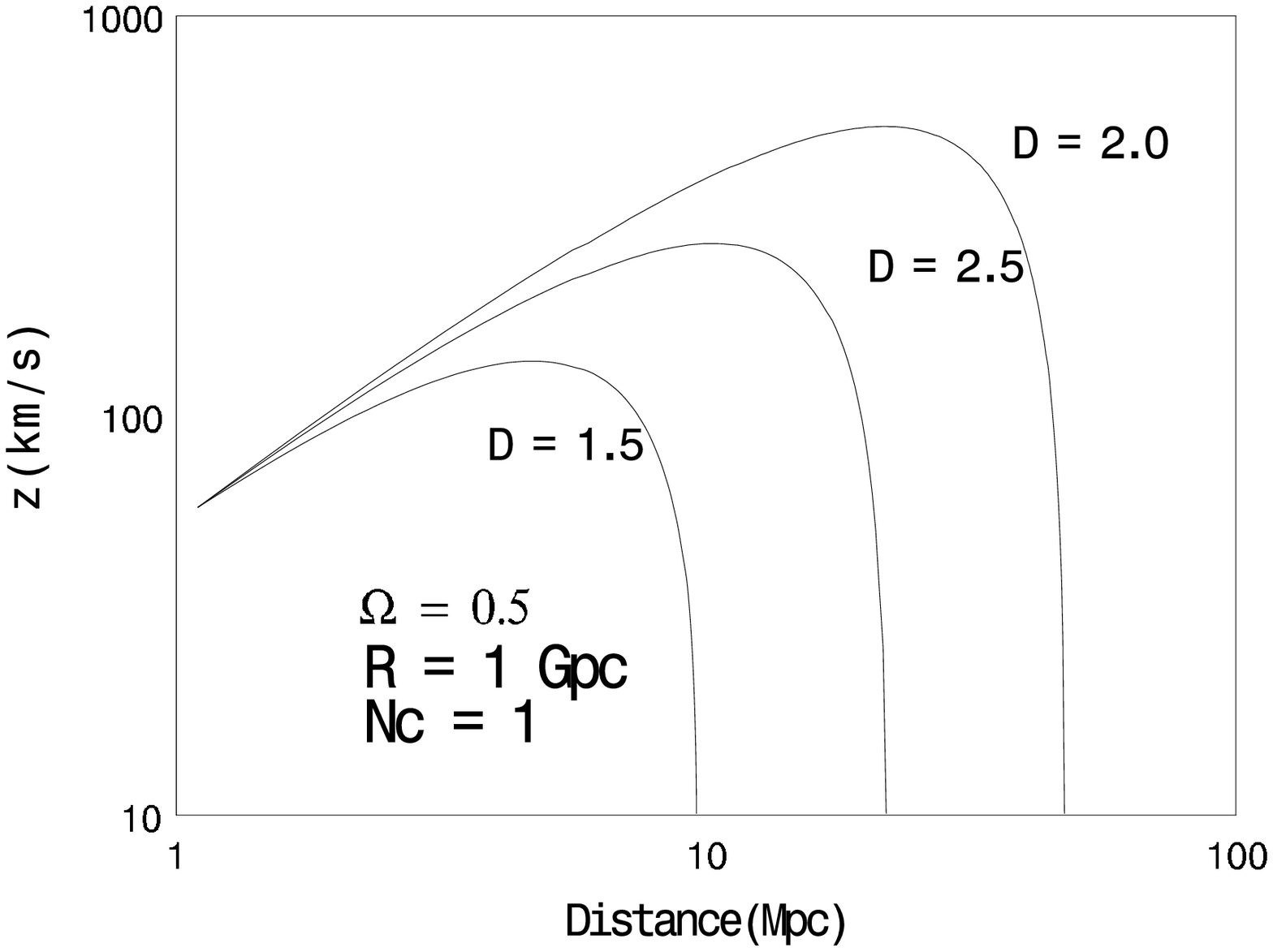}} \\
\epsfxsize= 5truecm{\epsfbox{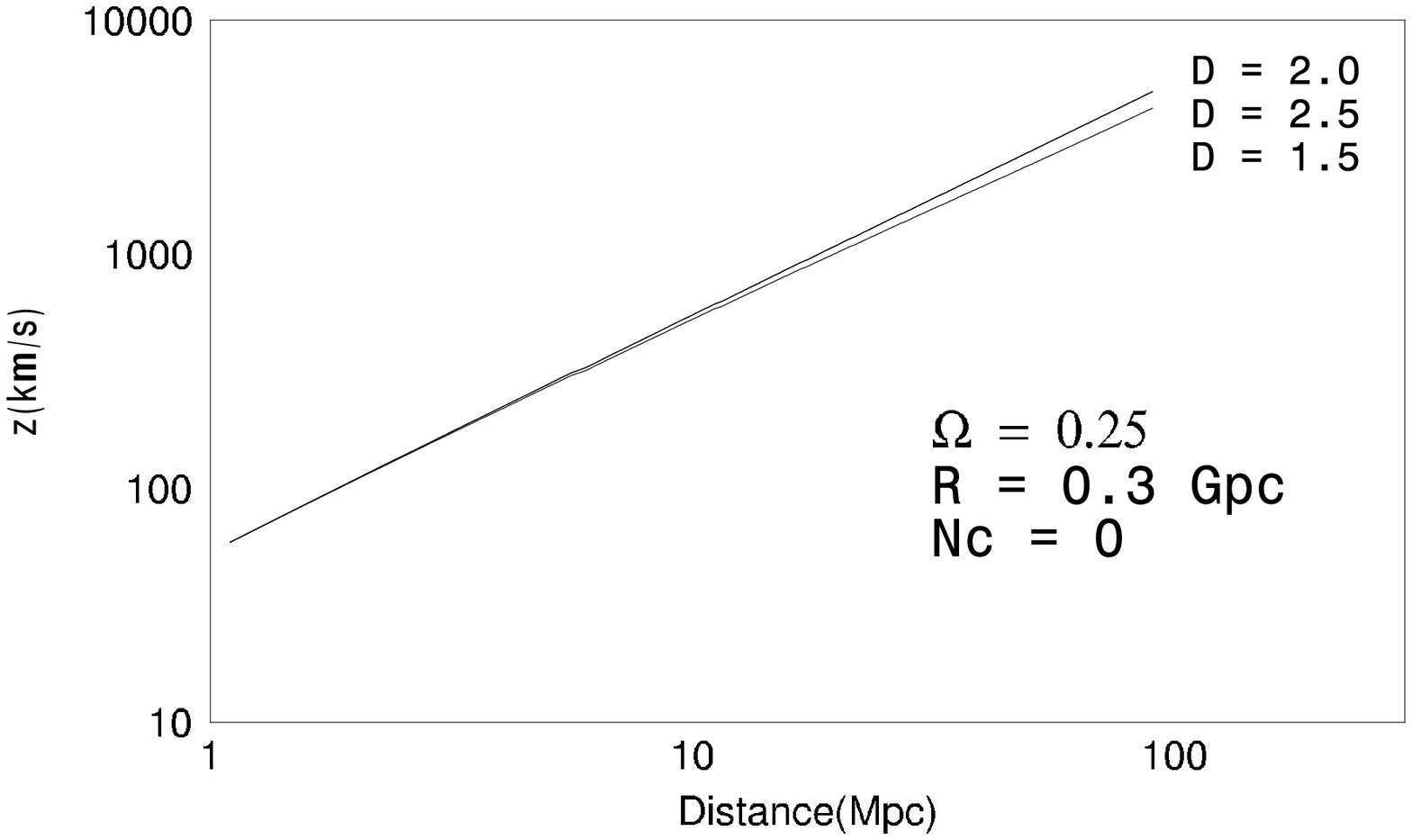}} & 
\epsfxsize= 5truecm{\epsfbox{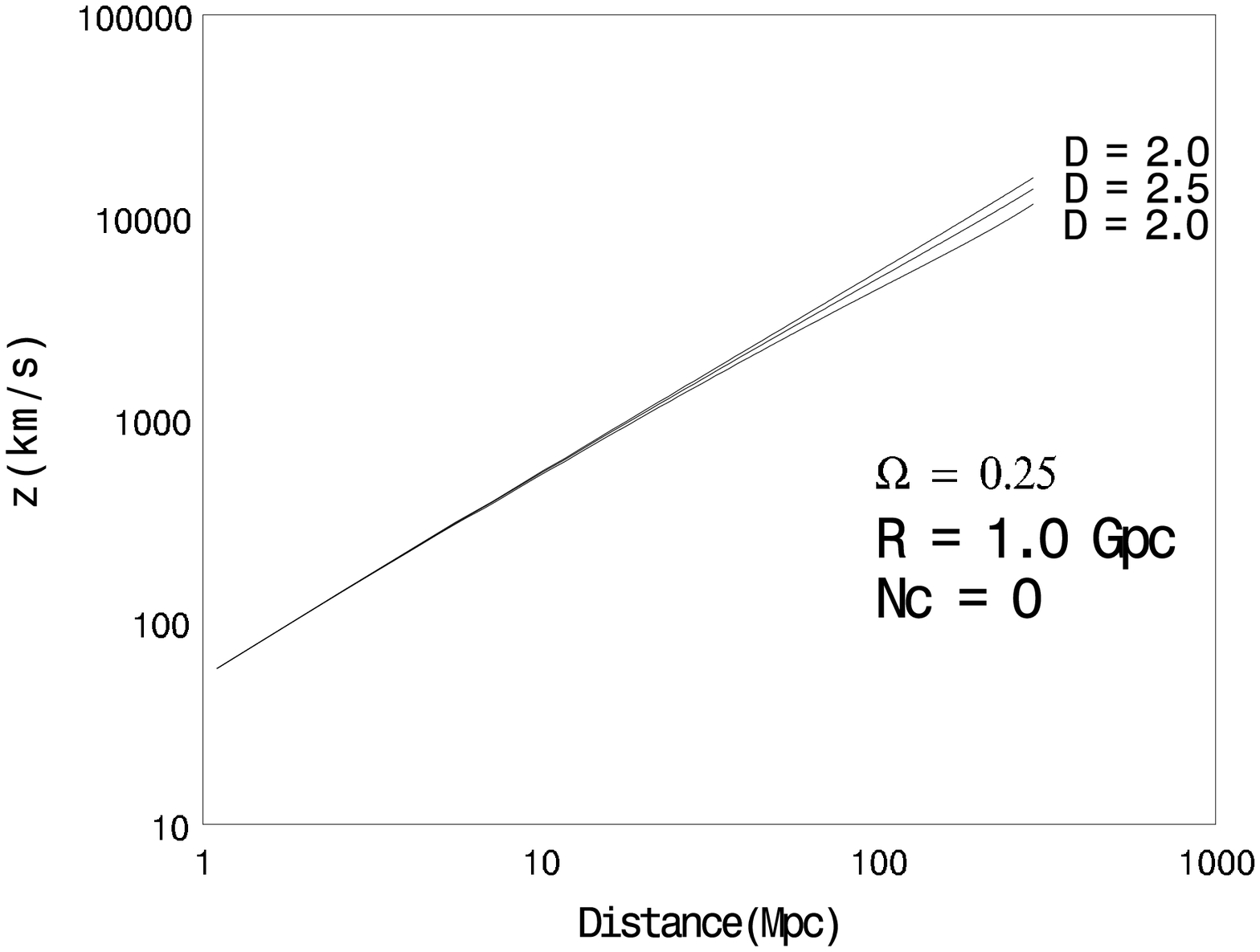}} & 
\epsfxsize= 5truecm{\epsfbox{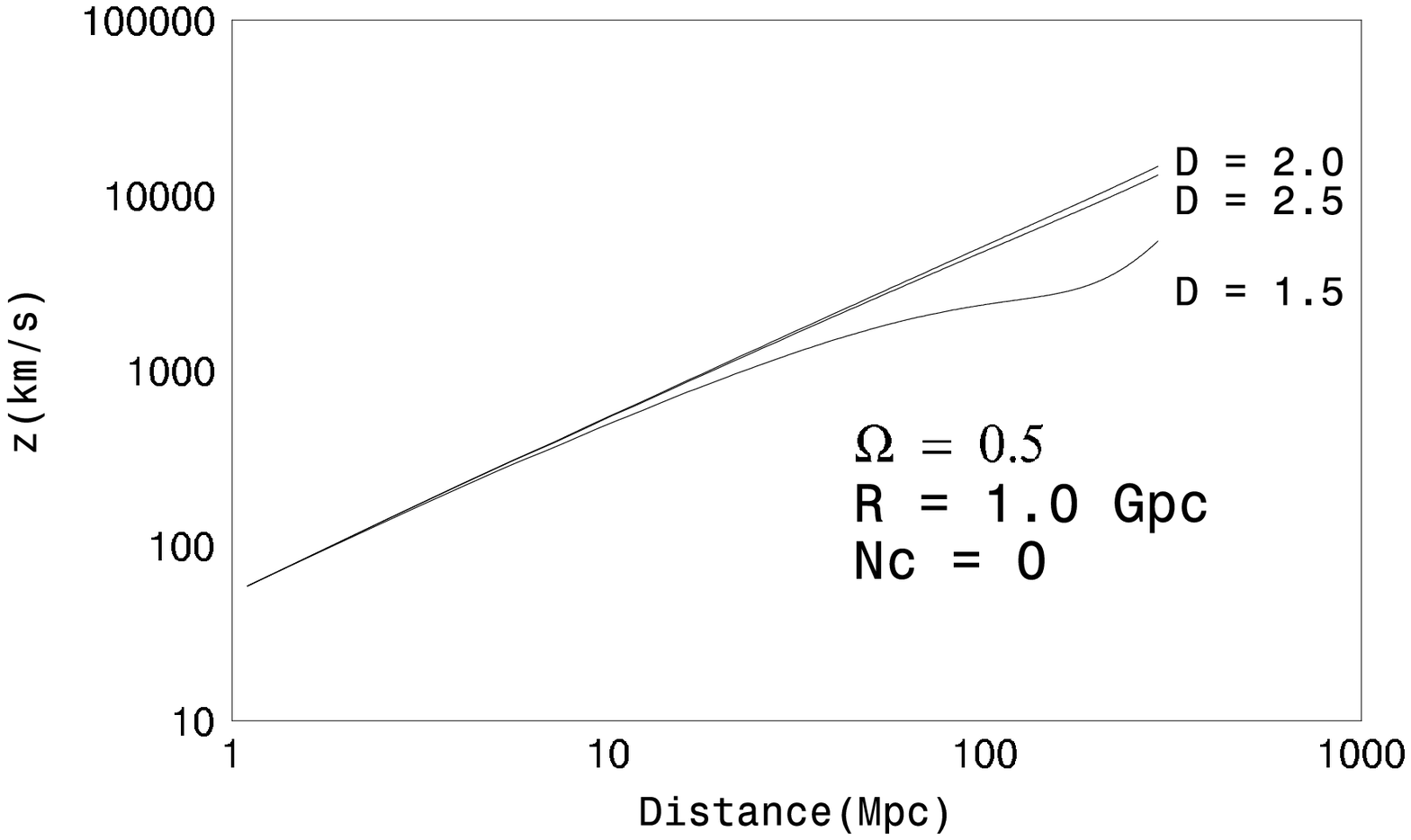}} 
\end{array}
$
\end{center}
\caption{
{\it
The graphs of velocity versus distance}
}
\end{figure}  

\end{document}